\documentclass[12pt]{article}
\usepackage[pctex32]{graphics}
\begin{document}
~
~
\vspace{1cm}
\begin{center} {\Large \bf  Tubular Solutions in NS5-brane, Dp-brane and Macroscopic Strings Backgrounds}
                                                  
\vspace{1cm}

                      Wung-Hong Huang\\
                       Department of Physics\\
                       National Cheng Kung University\\
                       Tainan, Taiwan\\

\end{center}
\vspace{1cm}
\begin{center} {\large \bf  Abstract} \end{center}
We present our searches for the tubular bound states of a D2-brane with $m$ D0-branes and $n$ fundamental strings in the NS5-brane, Dp-brane or macroscopic strings background by solving the Dirac-Born-Infeld equation.  The geometry of tubular configurations we considered  are taken to be parallel to the background branes or macroscopic strings.   The $n$ fundamental strings therein may be the circular strings $F_c$ or the straight strings $F_s$, which are along the cross section or the axis of the tube respectively.  We show that the stable tubular bound states of ($nF_s$, $m$D0, D2) which are prevented form collapse by the angular momentum, as original found in flat space, could be formed in the NS5-brane, D6-brane and macroscopic strings background.   However, they becomes unstable in the D2-brane and D4-brane background.   We also show that there are stable tubular bound states of ($nF_c$, $m$D0, D2)  and ($m$D0, D2), which are prevented form collapse by the gravitational forces in the backgrounds of D6-brane.  We discuss the properties of these tubular solutions and see that only that in the macroscopic strings background is a supersymmetric BPS configuration.

\vspace{2cm}
\begin{flushleft}
E-mail:  whhwung@mail.ncku.edu.tw\\
\end{flushleft}
%%%%%%%%%%%%%%%%%%%%%%%%%%%%%%%
\newpage
\section {Introduction}
A bunch of straight strings with D0-brane can be blown-up to a supersymmetric tubular D2-brane which is supported against collapse by the angular momentum generated by the Born-Infeld (BI) electric and magnetic fields [1-2].  The energy of  the tube is typical of 1/4 supersymmetric configurations, and a calculation confirms that the D2-brane configuration just describes preserves 1/4 supersymmetry, hence the name `supertube'.  Besides their intrinsic interest, supertubes are beginning to play an important role in black hole physics [3].  For example, the quantum states of the supertube counted by directly quantizing the linearized Born-Infeld action near the round tube or the geometrically allowed microstates with fixed conserved charges are shown to be in one-to-one correspondence with some black holes.

   In previous papers [4,5]  we have shown that a bunch of circular fundamental strings with D0-brane could be bound with D2-brane to form a stable tubular configuration which is prevented from collapsing by the magnetic force in the Melvin background or the forces in the curved D6-brane background.

  Recently  there are  interesting in investigating the dynamics of D-brane in the nontrivial background [6-8].  In [6] Kutasov found that the dynamics of the radial mode of the BPS D-brane in the background of NS5-brane resembles the tachyon rolling dynamics of unstable D-brane.   It was shown that the radion effective action takes exactly the same functional form as the tachyon effective action for unstable D-brane.  Therefore we can view the radion rolling dynamics as a sort of ``geometrical'' realization of tachyon rolling dynamics on unstable D-brane [9].  The interesting property stimulates many authors to study the subject in [7,8].  In [10] Panigrahi found that  the effective field theory description of the dynamics of the  BPS Dp-brane in the background of $N$ Dk-brane can be still mapped to the  tachyon like effective action for unstable Dp-brane. In [11] Burgess et. al. investigated the properties of a probe Dp$'$-brane in the Dp-brane background.  In [12] Bak et. al. studied the dynamics  of (p,q) string near macroscopic fundamental strings. 

In this paper we will extend the analysis performed in [5] to find the all possible 
tubular bound states of a D2-brane with $m$ D0-branes and $n$ fundamental strings in the nontrivial backgrounds of  NS5-brane, Dp-brane or macroscopic strings by solving the Dirac-Born-Infeld equation.  The geometry of tubular configurations we considered  are taken to be parallel to the background branes or macroscopic strings.   The $n$ fundamental strings therein may be the circular strings $F_c$ or the straight strings $F_s$, which are along the cross section or the axis of the tube respectively.  When the $n$ fundamental strings are along the cross, the circular F-strings are fusing inside the D2 worldsheet by converting itself into homogenous electric flux.  As the direct along the electric field is a circle with radius $R$ the open strings now stretch around the circle and the two ends join to each other with a finite probability [13-15].   In this case the stable tube is supported against collapse by the forces in the background.   On the other hand, when the fundamental strings (and thus the BI electric fields) are chosen to be along the axis of the tube then the stable tube is supported against collapse by the angular momentum generated by the Born-Infeld (BI) electric and magnetic fields, as that in [1-2].

  In section II we describe the Dirac-Born-Infeld action and set up our algorithm of finding the tube solution.  In section III we search the tube in NS5-brane background [16].  We show that  there are tubular bound states of  ($nF_s$, $m$D0, D2) with finite radius. However, the tubular bound states of ($nF_c$, $m$D0, D2) are unstable and will collapse into zero radius.

  In section IV we search the tube in Dp-brane background [17].  We show that the possible stable tubes are the bound states of ($m$D0, D2), ($n$$F_c$, $m$D0, D2), and ($n$$F_s$, $m$D0, D2) which are in the D6-brane background.   Especially, we also show that when a BPS ($n$$F_s$, $m$D0, D2)-tubes, which is supported against collapse by the angular momentum in flat space, is put into the D2-brane or D4-brane curved background, it will become metastable or unstable.   We give an argument to explain is property.

   In section V we search the tube in the macroscopic strings background [18]. We show that only the consititues $nF_s$, $m$D0 with  D2 could be formed as a stable tubular bound state and there does not exist stable ($nF_c$, $m$D0, D2)-tube.  

  Our results show that only in the macroscopic strings backgrounds could the stable tubular solution of ($n$$F_s$, $m$D0, D2) is a supersymmetric  BPS state.  The supersymmetry of the supertubes will be broken if they are put in the D6 or NS5 background.  We make a conclusion in the last section.

%%%%%%%%%%%%%%
\section {Dirac-Born-Infeld Action  in Curved Background}
  For convenience, we will use the units of string length and string coupling $l_s = g_s = 1$. The Dirac-Born-Infeld Lagrangian  used to describe the tubular bound state of $n$ fundamental strings, $m$ D0 and a D2  is written as [1]
$${\cal S} =  - \int_{V_3} dt\, dz \, d\theta \,e^{-\phi}~\sqrt{- \det (g + {\cal F})}\, + \int_{V_3} \, P\left[\sum_s C^{(s)}\right]\,\wedge  e^{\cal F},  ~~~~ {\cal F} =  F + B^{NS}\, ,\eqno{(2.1)} $$
where $g$ is the induced worldvolume 3-metric, $F$ is the BI 2-form
field strength, $B^{NS}$ is the NS-NS two form potential, $\phi$ is the dilaton field and $C^{(s)}$ is the s-form RR potential.   $P[C^{(s)}]$ denotes the pullback of the spacetime tensor $C^{(s)}$ to the brane worldvolume.   Substituting the metric of the background and the relevant fields into (2.1) we can find the Lagrangian of a static tube with finite radius $R$. 

   To proceed we shall first describe the geometry of tubular configuration.  The tube we considered are taken to be parallel to the background branes or macroscopic strings.   Denoting the coordinate $(t, z, x,...)$ as that of  the worldvolume of the background branes or macroscopic strings then, as described in figure 1, the tube along the z axis  will share two common coordinate $(t, z)$ with the background matters.

\vspace{1cm}
\hfil\scalebox{1}{\includegraphics{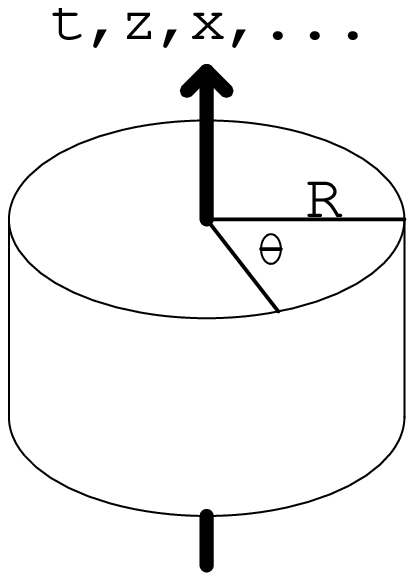}}\hfil\\
{\it ~~~Fig.1. The coordinate $(t, z, x, ..)$ is that of  the background branes or macroscopic strings and the coordinate $(t, z, \theta)$ is that of  the tube.  They have only two common coordinates $(t, z)$.}
\vspace{1cm}
\\
 We will allow a time-independent electric field $E$ and magnetic field $B$ such that the BI 2-form field strength is [1]
$$ {\bf F}= E \, dt\wedge dz + B \, dz\wedge d\theta ,~~   ~~~~~ \mbox{if  string  is straight along the axis of the tube}, \eqno{(2.2a)}$$
$$~{\bf F}= E \, dt\wedge d\theta + B \, dz\wedge d\theta,~~~\mbox{if string is circular along the cross section of tube}. \eqno{(2.2b)}$$
\\
Using the above definition of the worldvolume  coordinate and BI fields we can calculate the DBI action.  Then, we can define the momentum conjugate to the BI electric field $E$ [1]
$$\Pi \equiv {\partial{\cal L}\over \partial E}\, ,   \eqno{(2.3)}$$ 
and the corresponding Hamiltonian density of the tube is 
$$ {\cal H} \equiv \Pi E - {\cal L}. \eqno{(2.4)}$$
For an appropriate choice of units, the integrals 
$$m \equiv {1\over 2\pi} \oint d\theta \, B , ~~~~~~\mbox{and}~~~~~~n \equiv {1\over 2\pi}\oint d\theta \, \Pi \, , \eqno{(2.5)}$$
are, respectively, the conserved D0-brane charge and IIA string charge per unit length carried by the tube [1].  For the system with fixed charges we can therefore analyze the energy density relation (2.4) to see whether there is the stable tubular solution with finite radius. 

    We will in the following section use the above method to find the tube solutions in the various backgrounds.

%%%%%%%%%%%%%%%%%%%%%%%%%
\section {Tube in NS5-brane Background}
It is known that at weak string coupling $NS5$-branes are much heavier than D-branes - their tension goes like $1/g_s^2$, while that of D-branes goes like $1/g_s$. To study the dynamics of a tube in the background of fivebranes, we can therefore take the fivebranes to be static and study the motion of the tubular D2-brane under their gravitational potential. The background fields around $N$ $NS5$-brane are given by the CHS solution [16]. The metric, dilaton and NS-NS $B^{NS}$ field are

$$ds^2=dx_\mu dx^\mu + h_{NS}(x^n) dx^mdx^m ,$$
$$e^{2(\Phi-\Phi_0)}=h_{NS}(x^n)\, ,~~~~~~~~~~~~~~~~~$$
$$H_{mnp} = -\epsilon_{mnp}^q\partial_q\Phi\,.~~~~~~~~~~~~~~~~ \eqno{(3.1)}$$
Here $h_{NS}(x^n)$ is the harmonic function describing $N$ fivebranes, and $H_{mnp}$ is the field strength of the NS-NS $B^{NS}$ field. For the case of coincident fivebranes one has
$$h_{NS}(r) =1 + {N \over r^2} , \eqno{(3.2)}$$
where $r = |\vec x|$ is the radial coordinate away from the fivebranes in the transverse space labeled by $(x^6,\cdots, x^9)$.  

   Using the above metric and fields we will find the tube solutions in NS5-brane background by solving the associated DBI equation. 
%%%%%%%%%%%%%%%%%%%%%%%%%%%
\subsection {($nF_s$, $m$D0, D2)-tube in NS5-brane Background}
  The Lagrangian and corresponding Hamiltonian density  of the static ($nF_s$, $m$D0, D2)-tube with radius $R$ is
$$  L(R)  = -  \sqrt {(B^2 + h_{NS} \,R^2)\,h_{NS}^{-1} -  h_{NS} R^2 E^2}, ~~~~~~\eqno{(3.3)}$$

$$ {\cal H}(R) = {1\over\sqrt {R^2 + N}} \, \sqrt {(R^2 + B^2 + N)(R^2 +\Pi^2)}\, ,\eqno{(3.4)}$$
\\
respectively, in which $h_{NS}$ is defined by  (3.2) with $r \rightarrow R$.

   From (3.4) we have a simple relation 
 $$ {\cal H}'(R) =  {- R\over \sqrt{(R^2 + B^2 + N)(R^2 + \Pi ^2)(R^2 + N)^3}} \, \Big[B \sqrt{\Pi^2 - N} - (R^2 +N)\Big]\Big[B \sqrt{\Pi^2 - N} + (R^2 +N)\Big] .\eqno{(3.5)}$$
The solution of ${\cal H}'(R_c) = 0 $  found from the above equation represents a tube solution whose radius and energy density are
$$ R_c = \sqrt{B \sqrt{\Pi^2 - N} - N} \, , ~~~~~~~  {\cal H}(R_c) = B \sqrt{\Pi^2 - N}, \eqno{(3.6)}$$
respectively.  Thus, if the condition 
$$B \sqrt{\Pi^2 - N} - N > 0, \eqno{(3.7)}$$
is satisfied then we have a stable tube solution with finite radius.  Note that from (3.5) we know that 
 $${\cal H}'(R) \approx  {- R\over \sqrt{(B^2 + N)\, \Pi ^2\,N^3}} \, \Big[B \sqrt{\Pi^2 - N} - N)\Big]\Big[B \sqrt{\Pi^2 - N} + N)\Big] + O(R) < 0 ,\eqno{(3.8)}$$
$${\cal H}'(R) \approx 1 + O(1/R) > 0. ~~~~~\hspace{9cm}\eqno{(3.9)}$$
Thus (3.6) is indeed a stable tube solution with minimum energy density.   In figure 2 we plot the radius-dependence energy density (3.4) to see the property of the tube solution.

\vspace{1cm}
\hfil\scalebox{1}{\includegraphics{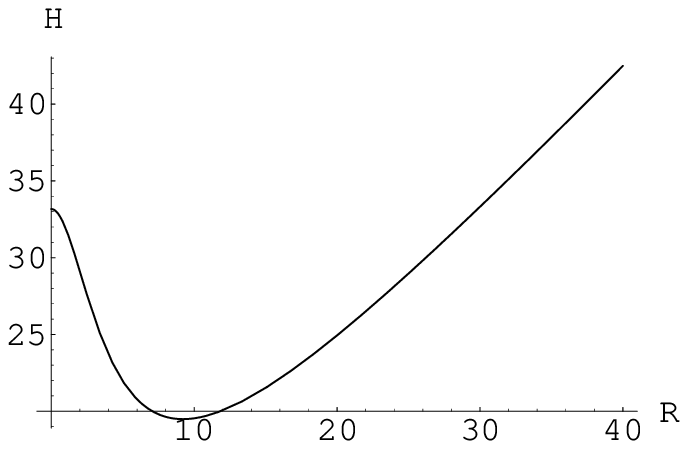}}\hfil\\
{\it ~~~Fig.2. The radius-dependence energy density (3.4) in the case of $N = \Pi =B =10$.}
\vspace{1cm}
\\
Figure 2 shows that the energy density at $R = 0$ for a supertube in an NS5-brane background is finite, as opposed to the case in flat space, in which it diverges.   This is because that the radius of the sphere transverse to the NS5-brane becomes constant in the near-horizon limit, i.e. for $R \rightarrow 0$. This  
makes the `centrifugal' energy finite in this limit. 

   The tube radius obtained in (3.6) reveals two interesting physical properties. 

  $\bullet$  First, the inequality 
$$R_c = \sqrt{B \sqrt{\Pi^2 - N} - N} <  \sqrt{B \Pi} =  R_c^{flat} \, ,\eqno{(3.10)}$$
in which $ R_c^{flat}$ is the tube radius in the flat space [1], tells us that the gravitational force in the background of NS5-brane will attract the tube configuration toward the $z$ axis and thus its radius will be smaller than that in the flat space. 

  $\bullet$ Second, to guarantee the reality of $R_c$ there must have sufficient number of  $nF_s$ and $m$D0, (for example $ \Pi^2 > N$)  to generate a sufficient angular momentum to support the tube against collapse in the curved NS5-brane background. 

\subsection {($nF_c$, $m$D0, D2)-tube in NS5-brane Background}
  The Lagrangian and corresponding Hamiltonian density  of the static ($nF_c$, $m$D0, D2)-tube with radius $R$ is
$$  L(R) = -  \sqrt {(B^2 + h_{NS} \,R^2)\,h_{NS}^{-1} -  h_{NS}^{-1} E^2}.~~~\eqno{(3.11)}$$

$$ ~~~{\cal H}(R) = \sqrt {(B^2 + N + R^2) \left(1 + \Pi^2 - {N \over R^2 + N}\right )}\, .\eqno{(3.12)}$$
\\
respectively, in which $h_{NS}$ is defined by  (3.2) with $r \rightarrow R$.  The energy density is an increasing functions of $R$ and we can therefore conclude that there does not exist ($nF_c$, $m$D0, D2)-tube in NS5-brane background.   

   Note that the classical tubular D2-D0-F1 bound states in NS5 background, which are extending transversely to NS5-branes have been found in [19]. 
%%%%%%%%%%%%%%%%%%%%%%%%%%%%%
\section {Tube in Dp-brane Background}
The metric, the dilaton $(\phi)$ and the RR field $(C)$ for a system of $N$ coincident $Dp$-brane are given by

$$ds^2=  h^{-{1\over 2}}_p ~\eta_{\alpha \beta}dx^\alpha dx^\beta + h^{1\over 2}_p ~\delta_{ij} dx^i dx^j, ~~~~~(\alpha, \beta = 0,..,p ; ~ i, j = p+1,..., 9), \eqno{(4.1)}$$
$$e^{2\phi}= h^{{3-p}\over 2}_p, ~~~~~~~ C_{0...p} = \,h^{-1}_p,~~~~~~~h_{p} = 1 + {N \over r^{7-p}}, \hspace{4.5cm}\eqno{(4.2)}$$
\\
where $h_p$ is the harmonic function of $N$ Dp-brane satisfying the
Green function equation in the transverse space [17]. 

   To proceed, we shall discuss the effects of the RR potential on the tube configuration. 

 1.  In our model we have a relation
$$\int_{V_3} \, P\left[\sum_s C^{(s)}\right]\,\wedge  e^{2\pi F} = \int_{V_3} \, \left[ P(C^{(3)}) + \,P (C^{(1)})\,\wedge  F\right].   \eqno{(4.3)}$$
Therefore, although the D2-brane background may provide a 3-form RR potential to give a nonzero contribution to the action, however, as described in figure 1, the coordinate $(t, z, x, ..)$, which is that of  the Dp-brane background, and $(t, z, \theta)$, which is that of the tube, have only two common coordinate $(t, z)$.  Thus the pullback of the form $C^{(3)}$ is always zero.  This shows a fact that the RR potential of the D2-brane background does not affect the tube configuration in our model. 

  2.  The Dp-branes are charged electrically and magnetically under RR field strengths.  The respective field strengths are
$$F_{t, x_1,...,x_p, r}^{(p+2)} = \partial_r h^{-1}_p, \eqno{(4.4)}$$
and 
$$F_{\theta_1, ..., \theta_{(8-p)}}^{(8-p)} = (7-p) (R_s)^{7-p}\, \sqrt{g^{(8-p)}}, \eqno{(4.5)}$$
in which $g^{(8-p)}$ is the metric of the unit $(8-p)$-sphere transverse to the Dp-brane, coordinates $\theta_1, ..., \theta_{8-p}$ are the spherical coordinates parameterizing the sphere and $R_s$ is the radius of the sphere.   Therefore, in the D6-brane background there is a one-form $A_\theta^{(1)}$ which will combine with two-form BI field strength $F_{t,z}$ to contribute the action, i.e. the second term in (4.3).  From (2.2) we see that the BI field strength on ($n$$F_s$, $m$D0, D2)-tubes has $F_{t,z}$ compoment, therefore the tube will feel this RR force under D6-brane background.  On the other hands, the ($n$$F_c$, $m$D0, D2)-tubes does not feel RR force.

  3.  From (4.5) we see that in D6-brane background there is a RR field strength  $F_{\theta_1,\theta_2} = R_s\, sin\theta_1$.   In a gauge in which the potential has only an $A_{\theta_2} =  R_s \, cos\theta_1$ \,component we can choose  the angular $\theta_2$ to be one of the tube coordinate $\theta$ (see figure 1). In this case the value of $R_s \, cos\theta_1$ will be equal to the tube radius $R$ and second term in (4.3) becomes $\int \,P (C^{(1)})\,\wedge F = 2\pi R E$. 

  4. As the tube we considered will along $z$ axis of the Dp-brane, as described in figure 1,  the possible values of  p are 2, 4, and 6.  These are the Dp-brane backgrounds to be considered below. 

   Using the above formula and after the calculations we have the following results:
\\
\\
(1) ($nF_s$, $m$D0, D2)-tube:
\\
$$ L =  - \, h_p^{p-2\over 4}\,\, R\, \,\sqrt { h_p^{-1}\left(1+{B^2\over R^2}\right) - E^2}+ 2\pi N\, E \delta_{p,6}.~~~~~~~~~~~~~~\eqno{(4.6)}$$
\\
$${\cal H} = \,\sqrt {h_p^{-1}\left(1 + {B^2\over R^2}\right) \left(R^2\, h_p^{p-2\over 2} + \tilde \Pi^2\right)}, ~~~~~\tilde \Pi \equiv  \Pi + 2\pi R\, \delta_{p,6}. \eqno{(4.7)} $$
\\
(2) ($nF_c$, $m$D0, D2)-tube:
\\
$$L = - \, h_p^{p-4\over 4}\,\sqrt {R^2 + B^2 - E^2}. ~~~~\eqno{(4.8)}$$
\\
$${\cal H} = \,\sqrt {\left(R^2 + B^2\right) \left( h_p^{p-4\over 2} + \Pi^2\right)}. \eqno{(4.9)} $$
\\
In the following subsections we will use the above formulas to find the all possible values of $p$ which allow the stable configurations of ($nF_i$, $m$D0, D2)-tube, in which $F_i$ = $F_s$ or $F_c$.

%%%%%%%%%%%%%%%%%%%%%%%%%%%
\subsection{($m$D0, D2)-tube solution in Dp-brane Background}
We search in this subsection some simple tube solutions with $n$ and/or $m$ = 0.

  Using  (4.7) and (4.9)  we have the relations:
$${\cal H}_{(D2)-tube} =  \left\{\begin{array}{ccc}
\sqrt{R^7\over R^5 + N } \, ,             & p = 2&       \\
R\,,                  & p = 4    & \\
\sqrt{R^2 +  R N }  \, ,                   & p = 6.   & \\
\end{array} \right . \eqno{(4.10)} $$

$${\cal H}_{(n  F_s , D2)-tube} = \left\{\begin{array}{ccc}
\sqrt{\left({R^5\over R^5 + N } \right)\left(R^2 + \Pi^2\right)} \, ,            & p = 2 ,&  \\
 \sqrt{R^2 + {\Pi^2 \,R^3\over R^3 + N }} \, , & p = 4 ,   & \\
\sqrt{R^2+  RN  + {\tilde \Pi^2  R \over R + N}}\, , & p = 6 .   & \\ 
\end{array} \right. \eqno{(4.11)} $$
\\
$${\cal H}_{(n  F_c , D2)-tube} =  \left\{\begin{array}{ccc}
\sqrt{R^2 \left({R^5\over R^5 + N } + \Pi^2\right)} \, ,  & p = 2 ,&  \\
\sqrt{R^2 \left(1+ \Pi^2\right)} \, ,           & p = 4 ,   & \\
\sqrt{R^2 \left({R + N \over R} + \Pi^2\right)} \, , & p = 6.   & \\ \end{array} \right. \eqno{(4.12)} $$
\\
The above energy densities are all the increasing functions of $R$ and therefore we conclude that there is no stable  (D2)-tube,  ($F_s$, D2)-tube nor  ($F_c$, D2)-tube on any Dp-brane background.

 We now turn to search the tube constructed by $m$D0 and D2.    Using (4.7) or (4.9) we have the relations:
$$~~{\cal H}_{(mD0 , D2)-tube} = h_p^{p-4\over 4} \sqrt{R^2 + B^2}= \left\{\begin{array}{ccc}
\sqrt{(R^2 + B^2)\,\left(R^5\over R^5 + N \right)} \, ,      & p = 2 ,&       \\
\sqrt{\left(R^2 + B^2\right)},           & p = 4 ,    & \\
\sqrt{\left(R^2 + B^2\right)\, \left(1 + { N \over R}\right)} \, ,                   & p = 6 .   & \\
\end{array} \right . ~~~~~\eqno{(4.13)} $$
The energy densities in the cases of p=2 and p=4 are the increasing functions of $R$ and thus there is no stable ($m$D0, D2)-tube in D2- or D4-brane background.  

    On the other hand, in the case of  p = 6 the energy density ${\cal H} \rightarrow \,\infty$   as R $\rightarrow$ 0 or $\infty $. This implies that there has a stable tube with a finite radius.  Thus we conclude that the $(mD0 , D2)$-tube can be stabilized by the forces from the background of D6 when the tube is parallel to the space of  D6 background. The tube  radius calculated from (4.13) is 

$$R_c = {N\over 6}\left[\left({\sqrt{27 B^2 - N^2}+ 3 \sqrt{3} B \over N} \right)^{2/3}+\left({\sqrt{27 B^2 - N^2} + 3 \sqrt{3} B \over N} \right)^{-2/3} -1\right]. \eqno{(4.14)} $$
\\
When $N \ll 1$ or  $N \gg 1$ we have a simple relation 
 $$R_c \rightarrow \left(N B^2/2\right)^{1/3}\, ~~~~~~~~~ as ~~~~N \ll 1.\eqno{(4.15a)} $$
$$R_c \rightarrow B - B^2/N\, ~~~~~~~~~ as ~~~~N \gg 1.\eqno{(4.15b)} $$
In figure 3 we plot the $N$-dependence of the tube radius $R_c$.   We see that if $N=0$ then there is no stable tube with finite radius, as is consistent with (4.15a).  However, if we  turn on the fields of background D6-brane then the forces therein could support the $(mD0 , D2)$-tube from collapse into zero radius.   Figure 3 shows that the tube radius is an increasing function of  $N$, the number of the D6-brane on the background geometry, as is consistent with (4.15).  This means that the background with larger value of $N$, which represents that there is the larger number of the D6-brane on the background geometry, will give a larger force to support the tube from collapse into zero radius.   Thus the radius of the tube will be getting larger.   This property clearly show that the ($m$D0, D2)-tube is stabilized by the forces from the D6-brane background. 

\vspace{1cm}
\hfil\scalebox{1}{\includegraphics{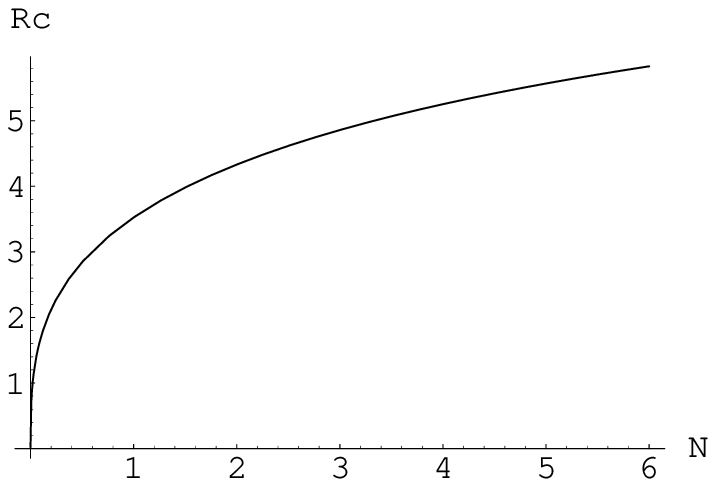}}\hfil\\
{\it ~~~Fig.3. $N$ dependence of  ($m$D0, D2)-tube radius $R_c$.   The tube is  in D6-brane background and has $B=10$.   The figure shows that the tube radius is an increasing function of  $N$.}
\vspace{1cm}

Note that in the case of $N \gg 1$ the background of   $N$ D6-branes could be regarded as a static geometry and we can study the motion of the tubular D2-brane under the D6-brane potential.

\subsection{($n$$F_c$, $m$D0, D2)-tube in Dp-brane Background}
In this subsection we search the tube of the bound state of $n$$F_c$, $m$D0 with D2.   In this case the tube is denoted by ($n$$F_c$, $m$D0, D2)-tube.  Using (4.9) we have the relations:
$$~~~{\cal H}_{(nF_c, mD0, D2)-tube} = \left\{\begin{array}{ccc}
\sqrt{\left(R^2  + B^2\right)\left({R^5\over R^5 + N } + \Pi^2\right)} \, ,            & p = 2 ,&  \\
\sqrt{\left(R^2  + B^2\right)\left(1+ \Pi^2\right)} \, ,           & p = 4 ,   & \\
\sqrt{\left(R^2  + B^2\right)\left({R + N \over R} + \Pi^2\right)} \, , & p = 6.   & \\ \end{array} \right. \eqno{(4.16)} $$
As same as the discussions below of (4.13) the above equation implies that only  in D6-brane background could we find a stable ($n$$F_c$, $m$D0, D2)-tube with finite radius.    The function form of tube  radius is to be complex to be cited.  However, when $N \ll 1$ or  $N \gg 1$ we have a simple relation 
 $$R_c \rightarrow \left({N B^2\over 2\, (1+\Pi^2)}\right)^{1/3}\, ~~~~~~~~~ as ~~~~N \ll 1.\eqno{(4.17a)} $$
$$R_c \rightarrow B - B^2 \, (1+\Pi^2)/N\, ~~~~~~~~~ as ~~~~N \gg 1.\eqno{(4.17b)} $$
The property of this tube solution is same as that of (4.13) and we conclude that the $(nF_c$, mD0 , D2)-tube can be stabilized by the forces in the background of D6 when the tube is parallel to the space of  D6 background.
%%%%%%%%%%%%%%%%%%%%
\subsection{($n$$F_s$, $m$D0, D2)-tube in Dp-brane Background}
In this  subsection we will search the stable tubes of the bound states of $n$$F_s$, $m$D0 with D2.  Note that, at first sight,  the angular momentum of these tubes, which is generated by the Born-Infeld (BI) electric and magnetic fields (i.e. $n$$F_s$ and $m$D0, as that in [1]), may support the tubes against collapse in Dp-brane background.   However we will see that it is not the case. 

 Using (4.7) we have the relation:
$$~~~{\cal H}_{(NFS_s, mD0, D2)-tube} = \left\{\begin{array}{ccc}
\sqrt{\left(R^2  + B^2\right)\left({R^3\over R^5 + N } \right)\left(R^2 + \Pi^2\right)} \, ,            & p = 2 ,&  \\
\sqrt{\left(R^2  + B^2\right)\left(1+ \Pi^2 \, {R\over R^3 + N }\right)} \, ,           & p = 4 ,   & \\
\sqrt{\left(R^2  + B^2\right)\left(1+ {N g_s l_s\over R} + \tilde \Pi^2 \, {1\over R(R + N )}\right)}\, , & p = 6 .   & \\ \end{array} \right. \eqno{(4.18)} $$
As same as the discussions below of (4.13) the above equation implies that only  in the D6-brane background could we find a stable ($n$$F_s$, $m$D0, D2)-tube with finite radius.   To clearly see the unstabilization of ($n$$F_s$, $m$D0, D2)-tube by the background D2- or D4-brane we plot  in figure 4 the radius dependence of the energy density of (4.18). 
\vspace{1cm}

\hfil\scalebox{1}{\includegraphics{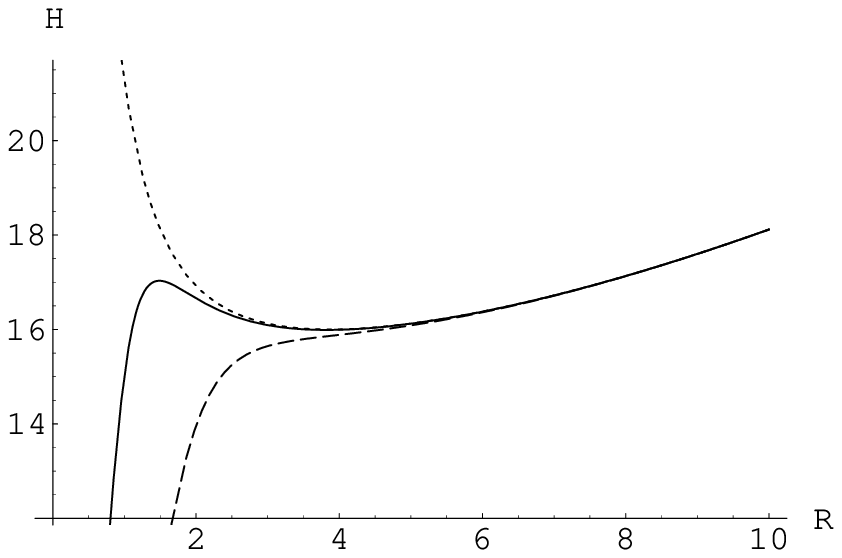}}\hfil\\
{\it ~~~Fig.4. Radius dependence of  the energy density ${\cal H}(R)$ in (4.18) for the case of p =2. The systems are taken to be $B=1$ while $N $ = 0, 1 and 15  in the dot line, solid line and dash line, respectively.  In the solid line, there is a finite radius at which the energy density becomes a local minimum.}
\vspace{1cm}

   From figure 4 we clearly see that increasing $N$ from zero to a finite value will shift the tube energy density from a global minimum (dot line in figure 4) to a local minimum (solid line in figure 4).   Thus the original supertube will now become a metastable configuration.   Once increasing $N$ to a large value the  the energy density ${\cal H}$ becomes an increasing function of radius $R$ (dash line in figure 4) and thus the metastable state disappears eventually.   These properties also appear in the case of p=4.  

We make the following comments to conclude this section

1.  In section 4.1 we have found that there is classically stable D2/D0 bound state in the D6 background.  This is because that the D6-branes' force on the D0-branes of the tube is repulsive (hence the divergence of the energy as $R$ goes to zero), whereas the tension of the tubular D2-brane provides a contracting force (hence the divergence of the energy as $R$ goes to infinity).    The ($m$D0, D2)-tube solution with finite radius could therefore be formed in the D6-brane background.

  2. The absence of stable supertubes in the D2 and D4 backgrounds is 
associated to the rate at which the radii of spheres transverse to the D2- and 
D4-branes decrease as $R \rightarrow 0$. 

%%%%%%%%%%%%%%%%%%%%%%%%
\section {Supertube in Macroscopic Strings Background}
The metric, the dilaton $(\phi)$ and the NS-NS $B^{NS}$ field  for a system of $N$ coincident macroscopic fundamental strings are given by 
$$ ds^2 = {1 \over h_f(r)} \Big( - dt^2 + dz^2 \Big) + dx^m dx^m , ~~~~~B^{NS}_{01} =  h_f(r)^{-1} - 1 , ~~~~e^{2 \phi} =   h_f(r)^{-1},   \eqno{(5.1)}$$
where $ r$ denotes spatial coordinates transverse to the macroscopic string, $r \equiv {\sum x^m x^m}$, and the harmonic function $h_f({r})$ solving the  transverse Laplace equation is [18]
$$ h_f(r) = \Big( 1 + {N \over r^6} \Big). \eqno{(5.2)}$$
Using the above metric and fields we will in below find the tube solutions in macroscopic strings background by solving the associated DBI equation. 
%%%%%%%%%%%%%%%%%%%%%%%%%%%
\subsection {($nF_s$, $m$D0, D2)-tube in Macroscopic Strings Background}
  The Lagrangian  of the static ($nF_s$, $m$D0, D2)-tube with radius $R$ is
$$  L = -  \sqrt {R^2\, h_f^{-1} + B^2 - (h_f^{-1} + E -1)^2\, R^2 \, h_f }.\eqno{(5.3)}$$
The displacement field is given by 
$$\Pi ={(h_f^{-1} + E -1)\, R^2 \, h_f\over \sqrt {R^2\, h_f^{-1} + B^2 - (h_f^{-1} + E -1)^2\, R^2 \, h_f }},\eqno{(5.4)}$$
in which $h_f$ is defined by (5.2) with $r \rightarrow R$.   The corresponding Hamiltonian density is 
$$ {\cal H}(R) = {1\over R^6 + N}\sqrt{(R^8+B^2 R^6+N B^2)(R^6+\Pi^2 R^4+N)} + {\Pi N\over R^6+N}.\eqno{(5.5)}$$
Using (5.4) we can form the relation ${\cal H'}(R_c) = 0$ find the following stable tube solutions.  
$$(I)~~~~R_c = \sqrt{B\Pi} \,, ~~~~~ {\cal H}(R) = B + \Pi ,\eqno{(5.6a)}$$
$$(II)~~~R_c = 0\, , ~~~~~~~~~~~ {\cal H}(R) = B + \Pi ,\eqno{(5.6b)}$$
The first solution has a finite radius and its energy form is nothing but the BPS conditions for 1/4-supersymmetric supertubes ought to obey.  In [12] Bak et. al. had found this solution and shown that this is indeed a supersymmetric tube.   In fact, the existence of supersymmetric (hence stable) supertubes in the F-string background was already shown in full detail in [1]. Supersymmetry follows from the fact that the supercharges preserved by the supertube are the same as those of an F1/D0 system. 

   In this paper, however, we have seen that the trivial solution with $R_c =0$ has the same energy density as that of supertube.     For clear, in figure 5 we plot the radius-dependence energy density (5.5) to see the relation between the two solutions.

\vspace{1cm}
\hfil\scalebox{1}{\includegraphics{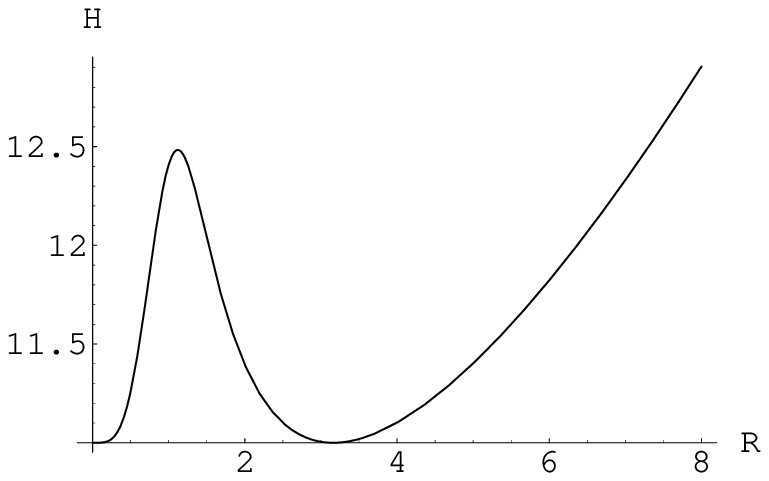}}\hfil\\
{\it ~~~Fig.5. The radius-dependence energy density (5.5) in the case of $N = B =10$ and $\Pi =1$.}
\vspace{1cm}
\\
Figure 5 shows that there are two ground states which could be formed once the constitutes $nF_s$, $m$D0 and D2  are put into the macroscopic strings background.  Thus, to find a tube with finite radius one shall initially put the constitutes outside the radius $R_c$ or larger then that is peak in figure 5. Otherwise,  the constitutes will be attracted toward the background macroscopic strings which are sitting at $R=0$, and become a trivial bound state.   However, as the potential barrier  is a finite value  the trivial bound state and non-trivial tubular bound state may be tunneling to each other quantum-mechanically.

  As the tube solution found in the other backgrounds, including the flat space, has a unique state of lowest energy the property shown in figure 5 is very special.  We belive that the property is the consequence of the distinguishing features of  the macroscopic strings background in which as one zooms into near horizon, ${r}\rightarrow 0$, both the string coupling $e^{\phi(r)}$ and the NS-NS $B^{NS}$ field decrease monotonically. However, it still remains to be clarified.

%%%%%%%%%%%%%%%%%%%%%%%%%%%
\subsection {($nF_c$, $m$D0, D2)-tube in Macroscopic Strings Background}
Before searching the  ($nF_c$, $m$D0, D2)-tube in macroscopic strings background there is an important issus shall be mentioned first.  

  Usually, we interpret the F-string bound states with D-brane as turning on electric field on D-brane, as the F-strings are fusing inside the D-brane worldsheet by converting itself into homogenous electric flux.  Thus the F-string on D-brane is characterized by the non-vanishing DBI electric field.  Using this property one attempts to conclude that the vanishing DBI electric field will imply no F-string.    This is, however, not the case in the macroscopic strings background in which there is a non-constant NS-NS $B^{NS}$ field.   In fact, from (5.4) we can easily see that only if $ E = 1- h_f^{-1} $ could the value $\Pi = 0$ and there is no F-string which is a straight string along the axis of the tube. (Note that the DBI electric field $E$ therein is along the z axis.)   This tells us that the DBI electric field of D-brane in the macroscopic strings background is not zero even if there is no F-string on it !

  With the above property in mind, the Lagrangian describing a static ($nF_c$, $m$D0, D2)-tube with radius $R$ is
$$  L = -  \sqrt {R^2\, h_f^{-1} + B^2 - E^2_\theta - (h_f^{-1} + E_z -1)^2\, R^2 \, h_f }.\eqno{(5.7)}$$
The momenta conjugate to $E_z$ and $E_\theta$ are 
$$\Pi_z = {h_f\, R^2 \, (h_f^{-1} + E_z -1)\over \sqrt {R^2\, h_f^{-1} + B^2 - E^2_\theta - (h_f^{-1} + E_z -1)^2\, R^2 \, h_f }},\eqno{(5.8)}$$
$$\Pi = {E_\theta\over \sqrt {R^2\, h_f^{-1} + B^2 - E^2_\theta - (h_f^{-1} + E_z -1)^2\, R^2 \, h_f }}.\eqno{(5.9)}$$
respectively.  We now set  $\Pi_z = 0$, i.e. letting $E_z = 1- h_f^{-1}$, then there will have no F-string along $z$ axis.  In this case,  the circular F-strings which are along the cross section of the tube are fusing inside the D-brane worldsheet by converting itself into homogenous electric flux $E_\theta$ and $\Pi$ corresponds to the charges of the  circular F-strings.   The energy density becomes 
$${\cal H} = \sqrt{1 + \Pi^2}\, \sqrt {R^2 \left(1- {N\over R^6+N}\right) + B^2}.\eqno{(5.10)}$$
As the energy density is an increasing functions of $R$ we thus conclude that there is no stable ($nF_c$, $m$D0, D2)-tube in macroscopic strings background.
%%%%%%%%%%%%%%%%%%%%%%%
\section{Conclusion}
The supertubes first found in [1] are the bunch of strings with D0-brane which are blown-up to a tubular D2-brane.   They are supported against collapse by the angular momentum.   In paper [4,5] we have shown that a bunch of circular fundamental strings with D0-brane could be bound with D2-brane to form a stable tubular configuration which is prevented from collapsing by the magnetic force in the Melvin background or the forces in the curved D6-brane background. 

In this paper we extend the analysis performed in [5] to find the all possible 
tubular bound states of a D2-brane with $m$ D0-branes and $n$ fundamental strings in the nontrivial backgrounds of  NS5-brane, Dp-brane or macroscopic strings by solving the Dirac-Born-Infeld equation.  The geometry of tubular configurations we considered are taken to be parallel to the background branes or macroscopic strings.   The $n$ fundamental strings therein may be the circular strings $F_c$ or the straight strings $F_s$, which are along the cross section or the axis of the tube respectively.  

  We show that there could have only the stable tubular bound state ($nF_s$, $m$D0, D2)  in the NS5-brane.   We explicitly see that the forces in the background of NS5-brane will attract the tube configuration toward the $z$ axis  and thus its radius will be smaller than that in the flat space.

   We also show that the forces in the Dp-brane curved background may support some tubular configurations of $n$ fundamental strings, $m$ D0, and D2 bound state form collapse.  Through the detailed analyses we have found that the possible stable tubes are the ($m$D0, D2)-tube, ($n$$F_c$, $m$D0, D2)-tube and ($n$$F_s$, $m$D0, D2)-tube which are in the D6-brane background.  We have also seen that, however, when a supersymmetric ($n$$F_s$, $m$D0, D2)-tube, which is supported against collapse by the angular momentum, is put into the D2-brane or D4-brane background, then the force coming from the background will make it unstable and render it collapse into the axis where the background branes is lying on. 

 We have also shown that there could have only the stable tubular bound state ($nF_s$, $m$D0, D2)  in the macroscopic strings background.  In this case the trivial bound state of $R=0$ has the same energy as that with finite radius.  As the potential barrier between them is a finite value the trivial bound state and non-trivial tubular bound state may be tunneling to each other quantum-mechanically.  We explain this special property as the consequence of the distinguishing features of  the macroscopic strings background in which as one zooms into near horizon, ${r}\rightarrow 0$, both the string coupling $e^{\phi(r)}$ and the NS-NS $B^{NS}$ field decrease monotonically. 

  Finally, among these tubular solutions we see that only that in the macroscopic strings background is a supersymmetric BPS configuration.   The supersymmetry of the supertubes will be broken if they are put in the D6 or NS5 background. 

\newpage
%%%%%%%%%%%%%%%%%%%%%%%
\begin{center} {\large \bf  References} \end{center}
\begin{enumerate}
\item D. Mateos and P. K. Townsend, ``Supertubes'', Phys. Rev. Lett. 87 (2001) 011602 [hep-th/0103030];\\
 R. Emparan, D. Mateos and P. K. Townsend, ``Supergravity Supertubes'', JHEP 0107 (2001) 011 [hep-th/0106012];\\
 D.~Mateos, S.~Ng and P.~K.~Townsend, ``Tachyons, supertubes and brane/anti-brane systems'', JHEP  0203 (2002) 016 [hep-th/0112054];\\
 P. K. Townsend, ``Surprises with Angular Momentum'', Annales Henri Poincare 4 (2003) S183 [hep-th/0211008];\\
 M. Kruczenski, R. C. Myers, A. W. Peet, and D. J. Winters,``Aspects of supertubes'',  JHEP 0205 (2002) 017 [hep-th/0204103].
\item D. Bak, K. M. Lee, ``Noncommutative Supersymmetric Tubes'',  Phys. Lett. B509 (2001) 168 [hep-th/0103148];\\
D. Bak and S. W. Kim, ``Junction of Supersymmetric Tubes,'' Nucl. Phys.  B622 (2002) 95 [hep-th/0108207];\\
 D. Bak and A. Karch, ``Supersymmetric Brane-Antibrane Configurations,'' Nucl. Phys. B626 (2002) 165 [hep-th/011039];\\
 D. Bak and N. Ohta, ``Supersymmetric D2-anti-D2 String,'' Phys. Lett.  B527 (2002) 131 [hep-th/0112034];\\
Y. Hyakutake and N. Ohta, ``Supertubes and Supercurves from M-Ribbons,'' Phys. Lett. B539  (2002) 153 [hep-th/0204161]; \\
N. E. Grandi and A. R. Lugo, ``Supertubes and Special Holonomy'', Phys. Lett. B553 (2003) 293 [hep-th/0212159].
\item  O. Lunin and S. D. Mathur  , ``AdS/CFT duality and the black hole information paradox'', Nucl.Phys. B623 (2002) 342 [hep-th/0109154];\\
S. D. Mathur, A. Saxena and Y. K. Srivastava, ``Constructing 'hair' for the three charge hole,'' Nucl.\ Phys.\ B \ 680 (2004) 415 [hep-th/0311092];\\
 I. Bena and P. Kraus, ``Three Charge Supertubes and Black Hole Hair'' , Phys.Rev. D70 (2004) 046003 [hep-th/0402144];\\
B. C. Palmer and D. Marolf , `` Counting Supertubes'',  JHEP 0406 (2004) 028 [hep-th/0403025];\\
D. Bak, Y. Hyakutake, S. Kim and N. Ohta, ``A Geometric Look on the Microstaties of Supertubes,'' [hep-th/0407253]; D. Bak, Y. Hyakutake, and N. Ohta, `` Phase Moduli Space of Supertubes,'' [hep-th/0404104];\\
Wung-Hong Huang, `` Entropy and Quantum States of Tachyon Supertube'', JHEP 0412 (2004) 002 [hep-th/0410264].
\item Wung-Hong Huang, ``Tube of (Circular F, D0, D2) Bound States in Melvin  Background'' Phys. Lett. B599 (2004) 301 [hep-th/0407230].
\item Wung-Hong Huang, ``Tubular Solutions of Dirac-Born-Infeld Action on Dp-Brane Background'' Phys. Lett. B608 (2005) 244 [hep-th/0408213].
\item D.~Kutasov, ``D-brane dynamics  near NS5-brane,'' [hep-th/0405058]; ``A geometric interpretation of the open string tachyon,'' [hep-th/0408073].
\item Y.~Nakayama, Y.~Sugawara and H.~Takayanagi, ``Boundary states for the rolling D-brane in NS5 background,'' JHEP  0407(2004) 020 [hep-th/0406173];\\
D.~A.~Sahakyan, ``Comments on D-brane dynamics near NS5-brane,''
JHEP 0410 (2004) 008 (2004) [hep-th/0408070];\\
S.~Thomas and J.~Ward, ``D-brane dynamics and NS5 rings,'' [hep-th/0411130];\\
J.~Kluson, ``Non-BPS D-brane near NS5-brane,'' JHEP  0411 (2004) 013 [hep-th/0409298]; ``Non-BPS Dp-brane in the background of NS5-brane on transverse R**3 x S**1,'' [hep-th/0411014].
\item  A.~Ghodsi and A.~E.~Mosaffa ``D-brane Dynamics in RR Deformation of NS5-branes Background and Tachyon Cosmology'' [hep-th/0408015];\\
B.~Chen, M.~Li and B.~Sun, ``D-brane near NS5-brane:  With electromagnetic field,''  JHEP 0412 (2004) 057 [hep-th/0412022];\\
Y.~Nakayama, K.~L.~Panigrahi, S.~J.~Rey and H.~Takayanagi, ``Rolling down the throat  in NS5-brane background: The case of electrified D-brane,'' [hep-th/0412038].
\item A.~Sen, ``Time and tachyon,'' Int.  J.  Mod.  Phys.  A 18 (2003) 4869 [hep-th/0209122];  ``Tachyon matter,'' JHEP  0207  (2002) 065 [hep-th/0203265]; ``Rolling tachyon,'' JHEP  0204 (2002) 048 [hep-th/0203211];``Tachyon dynamics 
in open string theory,'' [hep-th/0410103].
\item K.~L.~Panigrahi, ``D-brane dynamics in Dp-brane background,'' Phys. Lett. B  601 (2004) 64 (2004) [hep-th/0407134];\\
O.~Saremi, L.~Kofman and A.~W.~Peet, ``Folding branes,'' [hep-th/0409092];\\
J.~Kluson, ``Non-BPS Dp-brane in Dk-Brane Background,'' [hep-th/0501010].
\item C.~P.~Burgess, P.~Martineau,  F.~Quevedo and R.~Rabadan, ``Branonium,'' JHEP 0306 (2003) 037 [hep-th/0303170];\\
C.~P.~Burgess, N.~E.~Grandi,  F.~Quevedo and R.~Rabadan, ``D-brane chemistry,'' JHEP 0401 (2004) 067 [hep-th/0310010].
\item D.~Bak, S.~J.~Rey and H.~U.~Yee, ``Exactly soluble dynamics of (p,q) string near macroscopic fundamental strings,'' JHEP 0412 (2004) 008 [hep-th/0411099].
\item E. Witten, ``Bound States of Strings and p-brane,'' Nucl.\ Phys.\ B460 (1996) 335 [hep-th/9510135].
\item  C.G.~Callan, I.R.~Klebanov,  ``D-brane Boundary State Dynamics,'' Nucl. Phys.  B465 (1996) 473-486  [hep-th/9511173].
\item  N.~Seiberg, L.~Susskind and N.~Toumbas, ``Strings in background electric field, space/time non-commutativity and a new noncritical string theory,''
JHEP 0006 (2000) 021 (2000) [hep-th/0005040]; \\
I.R.~Klebanov and  J.~Maldacena, ``1+1 Dimensional NCOS and its U(N) Gauge Theory Dual'', I. J. Mod. Phys. A16 (2001) 922  [hep-th/0006085]; \\
U. H. Danielsson, A. Guijosa, ans M. Kruczenski, ``IIA/B, Wound and Wrapped,'' JHEP 0010 (2000) 020 [hep-th/0009182].
\item C.~G.~.~Callan, J.~A.~Harvey and A.~Strominger, ``Supersymmetric string solitons,'' [hep-th/9112030].
\item G.~W.~Gibbons and K.~Maeda, ``Black holes and membranes in higher dimensional theories with dilaton fields,'' Nucl.\ Phys.\ B298 (1988) 741.
\item A.~Dabholkar and J.~A.~Harvey, ``Nonrenormalization of the 
Superstring Tension,'' Phys.\ Rev.\ Lett.\  63 (1989 478 (1989); A.~Dabholkar, G.~W.~Gibbons, J.~A.~Harvey and F.~Ruiz Ruiz, ``Superstrings And Solitons,'' Nucl.\ Phys.\ B  340 (1990) 33.
\item Y. Hyakutake, ``Expanded Strings in the Background of NS5-branes via a M2-brane, a D2-brane and D0-branes,'' JHEP 0010 (2000) 020 hep-th/0112073].

\end{enumerate}
\end{document}